\documentclass{vgtc}                          % final (conference style)
\graphicspath{{figures/}{pictures/}{images/}{./}} % where to search for the images

\usepackage{times}                     % we use Times as the main font
\usepackage{tabu}                      % only used for the table example
\usepackage{booktabs}                  % only used for the table example
\usepackage{lipsum}                    % used to generate placeholder text
\usepackage{mwe}                       % used to generate placeholder figures

\usepackage{mathptmx}                  % use matching math font
\usepackage{enumitem}

\onlineid{0}

\vgtccategory{Research}

\vgtcinsertpkg

\title{Collaborative XRTactics: A Formative Study on Tactical Communication in Outdoor Team Sports}

\author{Ut Gong\thanks{e-mail: ugong@g.harvard.edu}\\ %
        \scriptsize Harvard John A. Paulson School of Engineering and Applied Sciences %
        \and Qihan Zhang\thanks{e-mail: Zzz7iHan@outlook.com}\\ %
     \scriptsize Glasgow School of Art %
        \and Ziqing Yin\thanks{e-mail: 
        zyin5@uw.edu}\\ %
     \scriptsize University of Washington %
        \and Stefanie Zollmann\thanks{e-mail: stefanie.zollmann@otago.ac.nz}\\ %
     \parbox{1.4in}{\scriptsize \centering University of Otago}}

\teaser{
  \centering
  \includegraphics[width=\linewidth]{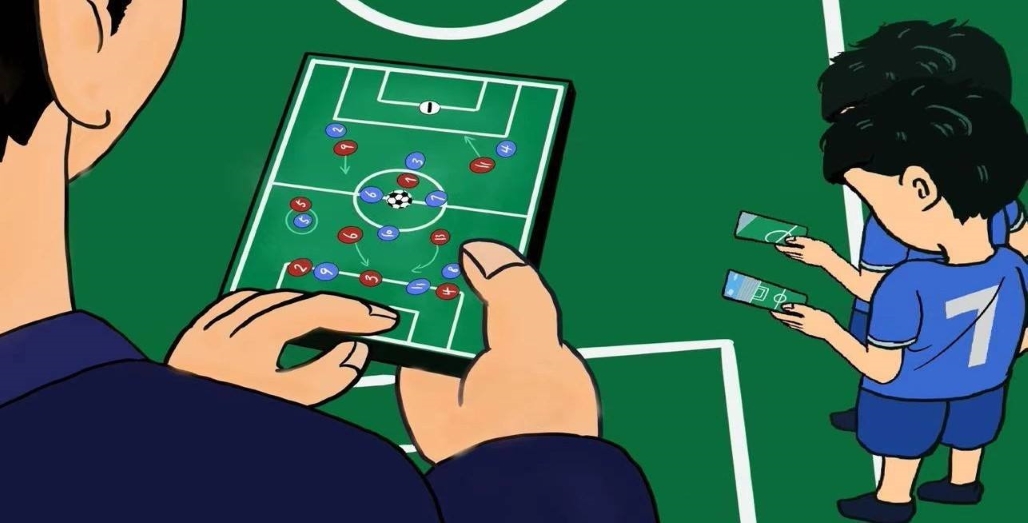}
  \caption{Exploring eXtended Reality (XR) solutions for enhancing tactical communication in outdoor team sports.}
  \label{fig:teaser}
}

\abstract{
    In team sports, effective tactical communication is crucial for success, particularly in the fast-paced and complex environment of outdoor athletics. This paper investigates the challenges faced in transmitting strategic plans to players and explores potential solutions using eXtended Reality (XR) technologies. We conducted a formative study involving interviews with 4 Division I professional soccer coaches, 4 professional players, 2 college club coaches, and 2 college club players, as well as a survey among 17 Division I players. The study identified key requirements for tactical communication tools, including the need for rapid communication, minimal disruption to game flow, reduced cognitive load, clear visualization for all players, and enhanced auditory clarity. Based on these insights, we propose a potential solution - a Mobile Augmented Reality (AR) system designed to address these challenges by providing real-time, intuitive tactical visualization and communication. The system aims to improve strategic planning and execution, thereby enhancing team performance and cohesion. This work represents a significant step towards integrating XR technologies into sports coaching, offering a modern and practical solution for real-time tactical communication. % filler text. Replace with your abstract.
} % end of abstract


\begin{document}

\firstsection{Introduction}

\maketitle

In team sports, effective collaboration is pivotal for success. Coaches and players alike rely heavily on strategic planning and real-time adjustments to outmaneuver their opponents. 
Traditionally, these strategies and tactics are visualized and communicated through various tools and methods, such as tactic boards, video analysis, and on-court illustration.
However, these methods have notable shortcomings, particularly when addressing the needs of outdoor sports like soccer. 
Understanding and improving these methods is an important topic because the effectiveness of tactical communication directly impacts team performance, player understanding, and overall team success. 
Enhancing these communication tools and methods can lead to significant advances in the way strategies are implemented and executed during games.

Tactic boards are the easiest and most common way coaches draw and communicate tactics during a game. 
However, in sports like soccer, where the field is large and involves many players, it is challenging for players to translate the 2D tactics drawn on the board into a 3D spatial understanding on the field. 
Additionally, the small size of the tactic board, typically held by the coach, requires players to gather closely to view it, which can limit effective communication. 
Video analysis offers a more dynamic solution, with state-of-the-art applications that facilitate easier annotation. However, this method also presents challenges. 
Coaches must invest time to annotate the videos, and typically, annotations are made on existing footage, which may not be as effective for real-time strategy adjustments. 
On-court demonstrations, while valuable for visual learning, cannot be used during practice or a game break, limiting their utility to practice sessions.
Current XR solutions, such as VIRD\cite{lin2024vird}, an immersive match video analysis for badminton coaching, are used for after-game tactic analysis and visualization. 
These systems provide immersive environments for analyzing player movements and strategies post-game. 
However, they do not support real-time annotation by coaches, making them unsuitable for in-game or real-time tactical adjustments. 
Addressing these challenges is crucial for advancing the effectiveness of team tactics in outdoor sports. 
Developing improved tools and methods for real-time collaboration can enhance a team's ability to adapt to the dynamic nature of the game, ultimately leading to better performance and more engaging gameplay for both players and spectators.

In this paper, we aim to offer design guidelines for outdoor team sports tactic discussions using eXtended Reality (XR). 
Our goal is to create a convenient environment for coaches and players to discuss and visualize team tactics more effectively. 
To this end, we followed a user-centered design approach during the study to address a key question: "What are the needs and challenges for coaches and players in demonstrating and understanding tactics?".
A formative user study was conducted in collaboration with professional coaches to address our question.
We interviewed 4 professional coaches from Division I teams, 4 professional players, 2 college club players, and 2 college club coaches. 
Based on these discussions, we identified key needs such as real-time tactical visualization, enhanced auditory communication, interactive and scalable solutions, and simplified cognitive load. 

In summary, our research seeks to bridge the gap between traditional tactical communication methods and the evolving needs of modern team sports. 
We aim to provide a more intuitive way for coaches and players to engage with tactical strategies, ultimately improving their performance and collaboration on the field.
\section{Related Work}

\subsection{Tactic Visualization}
Effective visualization of tactics is fundamental in team sports for strategic planning and in-game adjustments. Traditional methods include tactic boards, where coaches manually draw plays and formations, and video analysis, which involves annotating recorded game footage to highlight strategic elements. Traditional tactic boards,  offer simplicity and immediacy but are limited by their static nature and small size, making it challenging for players to translate 2D drawings into 3D spatial awareness on the field.

Advancements in digital technology have introduced more dynamic tools for tactic visualization. Interactive whiteboards and software applications like Tactic Board\cite{tacticalboard} and TacticalPad\cite{tacticalpad} allow for real-time manipulation of tactics and data analysis, enabling coaches to demonstrate and adjust strategies more fluidly. Video analysis tools have also evolved, incorporating features like motion tracking and interactive annotations, which provide deeper insights into player movements and team dynamics. For example, Bozyer et al. (2013) \cite{Bozyer2015} discussed the use of visual analytics to analyze and present football tactics using spatio-temporal data, highlighting the potential for advanced data analysis techniques to enhance tactical understanding. Burch et al. (2016)\cite{Perin2013} presented a comprehensive visual analysis approach to understanding soccer games, including tactics and strategies, emphasizing the importance of visual storytelling in sports analysis.

Despite these advancements, the challenge remains to find a method that combines the immediacy and simplicity of tactic boards with the dynamic, interactive capabilities of digital tools. The integration of real-time adaptability and ease of use is crucial for enhancing the tactical understanding and execution of players during high-pressure game situations. However, there are currently no real-time applications that can be seamlessly integrated into training sessions or used effectively during games, leaving a significant gap in the tools available for tactical communication.

\subsection{XR Tactic Visualization}
Extended Reality (XR), which includes Virtual Reality (VR) and Augmented Reality (AR), offers promising new avenues for tactic visualization in sports. XR technologies can create immersive and interactive environments that enhance the way tactics are communicated and understood. In VR, players and coaches can step into a virtual field where they can explore and manipulate tactical scenarios in a 3D space, providing a more intuitive understanding of spatial relationships and movements. Chen et al. (2023)\cite{chen2023settervision}introduced SetterVision, a VR-based tactical training system for volleyball players, which improves player setting skills and strategic thinking through immersive simulations. This approach underscores the potential of VR to improve tactical training and decision-making.

AR, on the other hand, overlays digital information onto the physical world, allowing for a blended experience where tactical instructions and visualizations can be seen in real-time on the actual field. This can be particularly useful for in-game adjustments, as it allows players to see tactical directions overlaied on their current environment without interrupting the flow of the game. 

Research has shown that XR technologies can improve the cognitive and perceptual understanding of complex tactical scenarios. Lin et al. (2024)\cite{lin2024vird} introduced VIRD, an immersive match video analysis system for high-performance badminton coaching, which provides an immersive environment for analyzing player movements and tactics, representing a step towards more interactive and intuitive tactical visualization tools. These technologies can facilitate better communication between coaches and players, enable more effective real-time strategy adjustments, and provide a deeper understanding of tactical concepts through immersive visualization.

However, the adoption of XR in sports is still in its nascent stages, with several challenges to address. These include ensuring that technology is user-friendly, accessible, and seamlessly integrated into the existing workflow of coaches and players. Additionally, the development of robust and responsive XR systems that can operate effectively in the dynamic and fast-paced environment of team sports remains a critical area of ongoing research.

In summary, while traditional methods of tactic visualization have their limitations, emerging XR technologies offer innovative solutions that can enhance tactical communication and understanding in team sports. Further research and development are needed to fully realize the potential of these technologies and integrate them effectively into the sports domain.
\section{Tactic Communication Challenges and Requirements in Outdoor Team Sports}

Effective communication is vital for success in any team sport, and it is particularly crucial in the dynamic environment of outdoor athletics. However, transmitting strategic plans to players often proves challenging. 
To explore these hurdles and identify potential solutions, we conducted a series of structured interviews with a diverse group of soccer athletes and coaches. 
This group included 4 professional coaches from Division I teams, 4 professional players, 2 college club coaches, and 2 college club players.

The interviews were conducted in a semi-structured format, allowing for both guided questions and open-ended discussions. 
This approach enabled us to gather detailed insights while also providing the flexibility to explore issues raised by the participants.
Each interview began with a set of core questions designed to assess the current methods of tactical communication used within the teams, including the use of tactic boards, video analysis, and verbal instructions. 
We then delved into the challenges faced in effectively conveying these strategies to players, particularly during outdoor situations such as games and crucial practice sessions.

Participants were encouraged to share their experiences and perspectives on the efficacy of different communication tools and methods. 
This included discussing the strengths and limitations of traditional methods like whiteboards and desktop software, as well as more modern technologies such as wearable AR and VR devices. 
The interviews also explored the potential for new technologies, to improve the clarity and speed of tactical communication.

Based on these discussions, the following key requirements were identified by coaches and players:

\begin{enumerate}[label=\textbf{R\arabic*}, left=0pt..2em]
    \item \textbf{Rapid Tactical Communication:}\label{req:R1} Coaches emphasized the need for solutions that enable quick and intuitive communication of strategies during dynamic practice sessions and critical game moments. 
    They highlighted that traditional methods, such as desktop software and whiteboards, are too time-consuming and often fail to capture the fluidity and complexity of outdoor team sports, leaving players struggling to grasp nuanced tactical concepts in real-time.

    \item \textbf{Minimizing Disruption to Game and Training Flow:}\label{req:R2} Both coaches and players highlighted the impracticality of using bulky wearable devices like Apple Vision Pro, Oculus Quest, or HoloLens during active gameplay or training. 
    They noted that these devices can be time-consuming to put on and adjust, which disrupts the flow of practice or a game. 
    Additionally, the discomfort of wearing such devices, especially when sweating, was a concern. 
    There is a strong preference for solutions that do not interrupt the natural flow of play and can be quickly and easily accessed.

    \item \textbf{Reducing Cognitive Load:}\label{req:R3} Players and coaches expressed the necessity for tools that simplify the translation of visual representations into actionable on-field decisions. 
    The current reliance on complex diagrams can overwhelm players, particularly in the heat of live gameplay, making it challenging to implement tactical instructions effectively.

    \item \textbf{Ensuring Clear Visualization for All Players:}\label{req:R4} A common concern was the limited scalability of existing solutions, which often rely on a single display for multiple players. 
    This setup can lead to visibility issues and hinder comprehension, especially for players positioned farther away or with obstructed views. 
    Coaches stressed the importance of tools that provide simultaneous, clear visualization for all team members, ensuring a shared understanding of strategic objectives.

    \item \textbf{Enhancing Auditory Clarity:}\label{req:R5} During games, the noise from crowds and on-field activities can impede players' ability to hear and understand the coach's instructions. 
    Coaches and players alike emphasized the need for communication tools that enhance auditory clarity, ensuring that all players can clearly hear strategic directions despite the ambient noise.

\end{enumerate}

Following the interviews, we conducted a survey among 17 Division I college soccer team players to explore their preferences for various tactical communication methods, including traditional tactic boards, mobile AR, MR, and VR systems. 
The survey asked players which display method they would prefer if a new tactical communication system were developed. 
The results indicated a strong preference for mobile AR, with 12 players favoring this approach, while 5 players preferred traditional tactic boards. 
The lack of interest in MR and VR solutions was attributed to concerns about practicality and comfort during gameplay.

These findings underscore the urgent need for innovative communication tools that enhance tactical understanding and execution in outdoor team sports. 
By developing solutions that address these key requirements, we can facilitate more effective real-time strategy communication, ultimately leading to improved team performance and cohesion.

\section{Potential Solution}

Based on these insights, we are currently developing a Mobile AR system designed to meet the players' and coaches' needs. 
The system aims to provide real-time, intuitive tactical visualization and communication, enabling players to receive and interact with tactical information in a seamless and immersive manner. 
The project focuses on integrating a web-based platform for coaches to input and update tactical strategies and a mobile AR application for players to view and engage with these strategies from a real-time, first-person perspective. 
This setup not only enhances the spatial understanding of tactical plans but also ensures that all team members receive consistent and clear instructions, regardless of the game's dynamics or external noise.

To facilitate \textbf{R1}, we utilize a pad platform where coaches can draw and annotate tactics intuitively using devices like Apple Pencil. 
This digital interface allows for quick and intuitive updates, enabling coaches to convey strategic adjustments swiftly during both games and practice breaks. 
This feature is particularly advantageous for making real-time decisions and ensuring that tactical instructions are communicated efficiently.

To address \textbf{R2}, we use mobile devices and tablets, which offer a significant advantage over AR or VR headsets. 
Unlike the cumbersome process of donning headsets, athletes can quickly access their phones during breaks, allowing them to review tactical instructions seamlessly without disrupting the game's pace. 
For higher-level players, assistant coaches can pre-set their phones to display the corresponding player's view within the system, ensuring that players are immediately ready to receive the necessary tactical information. 
This setup maintains continuity, reduces downtime, and helps players stay focused and prepared.

This immersive approach simplifies the translation of tactical plans into actionable decisions, reducing the mental effort required to interpret complex diagrams or abstract strategies.
As a result, players can react more quickly and accurately to tactical adjustments.

To ensure \textbf{R4}, each player is provided with their own device. 
This setup allows players to view tactical information from their specific on-field perspective, avoiding the common issue of multiple players trying to view a single shared display. 
By having individualized access to tactical visuals, players can better understand their roles and responsibilities, which enhances overall team coordination.

Lastly, to enhance \textbf{R5}, we convert spoken instructions from coaches into text displayed on players' phones. 
This feature is especially useful in noisy environments, where verbal communication can be challenging. 

In summary, our Mobile AR system leverages the accessibility and immediacy of mobile technology to improve tactical communication in soccer. 
By addressing the key requirements identified through our research—rapid tactical communication, minimal disruption, clear visualization, reduced cognitive load, and enhanced auditory clarity—the system promises to enhance strategic planning and execution, ultimately leading to improved team performance and cohesion. 
This innovative project represents a significant step forward in sports technology, offering a practical tool for teams to optimize their tactical communication during games and practices.

\section{Conclusion}

In conclusion, the need for improved tactical communication in outdoor team sports is evident from the challenges identified through our interviews and surveys. 
The Mobile AR system we are developing is specifically designed to address these challenges by providing rapid tactical communication, minimizing disruption during games, reducing cognitive load, ensuring clear visualization for all players, and enhancing auditory clarity. 
This system offers a practical and innovative solution that leverages mobile technology to provide real-time, intuitive, and accessible tactical instructions. 
By addressing the key requirements identified, our system aims to enhance the overall performance and cohesion of teams, making it a valuable tool for coaches and players alike. 
As we continue to refine and implement this system, we anticipate that it will set a new standard for tactical communication in sports, offering a modern approach that meets the demands of today's competitive environments.

\bibliographystyle{abbrv-doi}

\bibliography{template}
\end{document}